\documentclass[twocolumn, 
nobibnotes, 
aps, 
superscriptaddress, 
amssymb, 
amsmath, 
reprint, 
pra
]{revtex4-1}

\usepackage{amssymb}
\usepackage{amsmath}
\usepackage{graphicx}
\usepackage[colorlinks=true,breaklinks=true,allcolors=blue]{hyperref}
\usepackage{url}
\usepackage{tikz}
\usepackage{color}
\usepackage{bbold}
\usepackage{nicefrac}

\usetikzlibrary{decorations.text}

\usetikzlibrary{shapes.symbols, shapes.arrows, calc, shapes}

\def\equationautorefname~#1\null{Equation~(#1)\null}

\newcommand{\bra}[1]{\left<#1\right|}
\newcommand{\ket}[1]{\left|#1\right>}
\newcommand{\tr}{\text{tr}}
\newcommand{\su}{\uparrow}
\newcommand{\sd}{\downarrow}
\renewcommand{\vec}[1]{\mathbf{#1}}
\newcommand{\svec}[1]{\boldsymbol{#1}}

\begin{document}

\title{Neural network quantum state tomography in a two-qubit experiment}

\author{Marcel Neugebauer}
\author{Laurin Fischer}
\email{lfischer@physi.uni-heidelberg.de}
\author{Alexander J\"ager}
\affiliation{Physikalisches Institut, Universit\"at Heidelberg, Im Neuenheimer Feld 226, 69120 Heidelberg, Germany}
\author{Stefanie Czischek}
\affiliation{Kirchhoff-Institut f\"ur Physik, Ruprecht-Karls-Universit\"at Heidelberg, Im Neuenheimer Feld 227, 69120 Heidelberg, Germany}
\author{Selim Jochim}
\author{Matthias Weidem\"uller}
\affiliation{Physikalisches Institut, Universit\"at Heidelberg, Im Neuenheimer Feld 226, 69120 Heidelberg, Germany}
\author{Martin G\"arttner}
\email{martin.gaerttner@kip.uni-heidelberg.de}
\affiliation{Kirchhoff-Institut f\"ur Physik, Ruprecht-Karls-Universit\"at Heidelberg, Im Neuenheimer Feld 227, 69120 Heidelberg, Germany}
\affiliation{Physikalisches Institut, Universit\"at Heidelberg, Im Neuenheimer Feld 226, 69120 Heidelberg, Germany}
\affiliation{Institut f\"ur Theoretische Physik, Ruprecht-Karls-Universit\"at Heidelberg, Philosophenweg 16, 69120 Heidelberg, Germany}

\date{\today}

\begin{abstract}
We study the performance of efficient quantum state tomography methods based on neural network quantum states using measured data from a two-photon experiment. 
Machine learning inspired variational methods provide a promising route towards scalable state characterization for quantum simulators. While the power of these methods has been demonstrated on synthetic data, applications to real experimental data remain scarce.
We benchmark and compare several such approaches by applying them to measured data from an experiment producing two-qubit entangled states.
We find that in the presence of experimental imperfections and noise, confining the variational manifold to physical states, i.e.\ to positive semi-definite density matrices, greatly improves the quality of the reconstructed states but renders the learning procedure more demanding. 
Including additional, possibly unjustified, constraints, such as assuming pure states, facilitates learning, but also biases the estimator.
\end{abstract}
%

\maketitle

\section{Introduction}

Quantum simulation experiments allow the investigation of quantum many-body problems beyond the regime accessible with classical computers, with potential applications in fundamental physics, quantum chemistry and materials science \cite{Georgescu2014}.
With quantum simulators reaching sizes of tens and soon hundreds of qubits, the question arises how one can validate such devices to confirm that they faithfully emulate the targeted model. 
Using full state tomography for characterizing the prepared quantum states is not practical beyond the few particle regime as the number of measurements required scales exponentially with the system size.
Thus, efficient methods for state characterization are needed, meaning that the required resources, namely the number of experimental repetitions and the post-processing time, should scale economically, i.e.\ sub-exponentially, with the system’s size. 
Several approaches have been proposed in recent years to solve this problem, including permutationally invariant tomography \cite{Toth2010, Moroder2012}, compressed sensing \cite{Gross2010, Riofrio2017}, and tomographic schemes based on tensor network states \cite{Cramer2010, Baumgratz2013, Lanyon2017}. 
The central idea underlying these methods is to make a variational ansatz for the prepared state and adjust the variational parameters to best fit the measured data. 
To be efficient, the number of variational parameters should scale polynomially in the system size. Consequently, one searches the best fitting state among a restricted set of quantum states. The justification for imposing such constraints is usually given by \emph{a priori} known physical properties, e.g.\ a small amount of entanglement in the case of tensor network states.

A more flexible class of efficient tomography methods, requiring less prior knowledge about the prepared state, uses a data-driven approach inspired by rapid advances in machine learning. These techniques, known as neural network quantum state (NQS) tomography \cite{Torlai2020}, use generative modeling, a strategy for unsupervised learning, to reconstruct unknown quantum states. NQS exploit the expressivity of neural network models for finding efficient representations of quantum states. They were introduced by Carleo and Troyer \cite{Carleo2017} and have subsequently been investigated intensively (see \cite{Carleo2019, Torlai2020, Carrasquilla2020} for recent reviews). 
When employed for quantum state tomography, the model parameters of an NQS are adjusted, or learned, to best fit a set of experimentally measured data. 
If one assumes that the experimentally prepared state is pure, the original NQS ansatz can be employed for this \cite{Torlai2018}. However, in any experimentally realistic scenario noise and coupling to an uncontrolled environment require the adoption of a mixed state description. Thus, the density matrix of the state has to be parameterized, which can be achieved either by the so-called latent state purification ansatz \cite{Torlai2018a} or by directly modeling the outcome probabilities of a tomographically complete measurement \cite{Carrasquilla2019}.
These methods are facing a number of conceptual and practical challenges, in particular when it comes to applying them to real experimental data \cite{Torlai2019, Palmieri2020, Cha2020}, where the assumption of independent and identically distributed (i.i.d.) experimental runs that underlies all quantum state tomography methods may not be fulfilled. In particular, the measurement itself is prone to errors and noise. 

Here we compare different methods for NQS tomography by applying them to real experimental data in the case of a two-qubit system.
Specifically, we use a parametric down conversion source producing entangled photon pairs.
In such a small system the fitted NQS typically over-parameterizes the state, such that the finite representational power of the employed variational ansatz does not limit the quality of the reconstruction.
Nevertheless, we find that the learning procedure can be cumbersome. 
We benchmark the employed NQS tomography methods with maximum likelihood estimation (MLE) using the full density matrix and with direct measurements of Bell correlations.
We evaluate the performance of these methods using measured as well as synthetic data, and for different choices of tomographically complete measurements. 
Our study will provide guidance when taking the next steps towards applying NQS tomography methods in regimes beyond the reach of traditional methods.

\section{NQS tomography methods}
\label{sec:methods}

In this section we briefly review the quantum state tomography methods that we employed. More details can be found in the original publications \cite{Torlai2018, Torlai2018a, Carrasquilla2019}.

The quantum state of a system is fully specified by its density matrix $\rho$, which is a Hermitian positive semi-definite matrix with unit trace. Without taking into account the trace constraint, this matrix is specified by $d^2$ real parameters, where $d$ is the dimension of the Hilbert space. To infer all these parameters uniquely, a tomographically complete measurement, i.e.\ a measurement with at least $d^2$ outcomes needs to be carried out. For a general measurement, each of the outcomes corresponds to a positive operator $M_\vec{a}$ such that the probability of outcome $\vec{a}$ is given by
\begin{equation}
\label{eq:Pa}
    P(\vec{a})=\tr[\rho M_\vec{a}]\,.
\end{equation}
The set of operators $\{M_\vec{a}\}_\vec{a}$ constitutes a so-called positive operator valued measure (POVM) and fulfills $\sum_\vec{a} M_\vec{a}=\mathbb{1}$ (corresponding to the overall normalization $\sum_\vec{a} P(\vec{a})=1$).
The density matrix can be reconstructed uniquely from the probabilities $P(\vec{a})$ if the number of constraints imposed by the measurements matches (or exceeds) the number of parameters of the density matrix. This approach to reconstruct the state $\rho$ by inverting the linear system in Eq.~\eqref{eq:Pa} is typically referred to as \emph{linear reconstruction} \cite{Carrasquilla2019}.
In traditional quantum state tomography, all outcome probabilities are measured and the full density matrix is reconstructed. 
Linear reconstruction has the drawback that for finite experimental statistics the reconstructed $\rho$ will in general not be positive semi-definite, i.e.\ it will not be a physical density matrix, which can lead to unphysical results such as negative state probabilities or fidelities exceeding unity. To avoid this, one can parameterize the density matrix in a way that ensures positivity and determine the physical state that most closely resembles the measured distribution by maximum likelihood estimation (MLE) \cite{Paris2004}. 
The scaling issues of this method become obvious when we consider, for example, the case of $N$ spin $1/2$ particles, where the number of required measurements is $d^2=4^N$, thus scaling exponentially in the system size.
In efficient quantum state tomography one tries to overcome this problem by finding a variational ansatz for parameterizing either $\rho$ or alternatively the probability distribution $P$ with a number of variational parameters that scales polynomially in the system size. These variational parameters are adapted to best match the observed measurement outcomes.
Specifically, NQS tomography uses variational approaches that are inspired by generative models used in machine learning.

Before describing the NQS variational approach in detail, we need to specify what type of tomographically complete measurements we use. We are concerned with qubit systems, i.e.\ arrays of $N$ two-level systems. In this case it is natural to assume that each qubit is read out individually and thus the POVM elements are product operators $M_\vec{a}=M^{(1)}_{a_1}\otimes\ldots\otimes M^{(1)}_{a_N}$. If the single-qubit POVMs $\{M^{(1)}_{a_i}\}_{a_i}$ are tomographically complete, then this is also the case for $\{M_\vec{a}\}_\vec{a}$. We implemented two different tomographically complete measurements corresponding to the following single-qubit POVMs. 
First, the Pauli  POVM, or Pauli-6 POVM, consists of the projectors 
\begin{equation}
\label{eq:pauli6}
    \{M_{a}^{(1)}\}_{a} = \bigcup_{\alpha=x,y,z}\left\{
    \frac{1}{3}\ket{\su_\alpha}\bra{\su_\alpha},
    \frac{1}{3}\ket{\sd_\alpha}\bra{\sd_\alpha}
    \right\}
\end{equation}
where $\ket{\su_\alpha}$ and $\ket{\sd_\alpha}$ are the eigenstates of the Pauli-operator $\sigma_\alpha$ with eigenvalue $\pm 1$. By summarizing the $-1$ outcomes into a single POVM element one obtains 
\begin{align}
\label{eq:pauli4}
    M^{(1)}_0 &= \frac{1}{3}\ket{\su_x}\bra{\su_x} \\
    M^{(1)}_1 &= \frac{1}{3}\ket{\su_y}\bra{\su_y} \\
    M^{(1)}_2 &= \frac{1}{3}\ket{\su_z}\bra{\su_z} \\
    M^{(1)}_3 &= \mathbb{1} - M_0-M_1-M_2 \,, \label{eq:pauli4_lastline}
\end{align}
which we call the Pauli-4 POVM.
The second measurement setting we consider is the so-called tetrahedral POVM, which consists of the sub-normalized projectors onto states pointing into the corners of a regular tetrahedron on the Bloch sphere, i.e.\ $M^{(1)}_{a}=
(\mathbb{1}+\boldsymbol{s}_a\cdot\boldsymbol{\sigma})/4$,
with
$\boldsymbol{s}_0=(0,0,1)$, $\boldsymbol{s}_1=(\nicefrac{2\sqrt{2}}{3},0,-\nicefrac{1}{3})$,
$\boldsymbol{s}_2=(-\nicefrac{\sqrt{2}}{3},\sqrt{\nicefrac{2}{3}},-\nicefrac{1}{3})$,
and $\boldsymbol{s}_3=(-\nicefrac{\sqrt{2}}{3},-\sqrt{\nicefrac{2}{3}},-\nicefrac{1}{3})$ and $\boldsymbol{\sigma}$ the vector of Pauli-operators.
The Pauli-4 and tetrahedral POVM have the minimal number of four possible outcomes for each qubit required for being tomographically complete. In this case the overlap matrix $T_{\vec{a}\vec{a}'}=\mathrm{Tr}\left[M_{\vec{a}}M_{\vec{a}'}\right]$ is invertible and density matrix can be reconstructed as 
\begin{equation}
\label{eq:lin_rec}
    \rho=\sum_{\left\{\vec{a}\right\}}P(\vec{a})\mathcal{Q}_{\vec{a}} \,,
\end{equation}
where the operators $\mathcal{Q}_{\vec{a}}$ are given by $\mathcal{Q}_{\vec{a}}=\sum_{\left\{\vec{a}'\right\}} T_{\vec{a}\vec{a}'}^{-1}M_{\vec{a}'}$. As the overlap matrix factorizes for product POVMs, the inversion can be done on the single-spin level.
Analogously to $\rho$, any Hermitian operator, representing an observable, can be decomposed into the complete set of operators $\{M_\vec{a}\}_\vec{a}$, which allows to directly extract its expectation value from the POVM probability distribution $P(\vec{a})$. 

The probability distribution $P(\vec{a})$ resulting from a tomographically complete measurement, can be approximated using generative modeling approaches from machine learning \cite{Carrasquilla2019}, which are known to be universal function approximators \cite{leroux2008}.
Once such a model has been trained to represent the measured distribution $P(\vec{a})$, expectation values of observables can be calculated efficiently by drawing samples from the model distribution.
Here we focus on a restricted Boltzmann machine (RBM) ansatz, which is a rather veteran model in machine learning but has proven large representational power making it suitable for encoding large classes of quantum many-body states \cite{melko2019restricted}.
The RBM ansatz function is
\begin{equation}
\label{eq:RBM}
    Q(\vec{a};\mathcal{W})=\frac{1}{Z\left(\mathcal{W}\right)}\sum_{\left\{\vec{h}\right\}}\mathrm{exp}\left[-E\left(\vec{a},\vec{h};\mathcal{W}\right)\right],
\end{equation}
which is the marginal over all hidden states $\vec{h}=(h_1,\ldots, h_M)$, $h_j\in \{0,1\}$, of the joint Boltzmann distribution $\mathrm{exp}\left[-E\left(\vec{a},\vec{h};\mathcal{W}\right)\right]/Z\left(\mathcal{W}\right)$.
The network energy $E\left(\vec{a},\vec{h};\mathcal{W}\right)=-\sum_{i,j}a_iW_{i,j}h_j-\sum_{i}a_id_i-\sum_{j}h_jb_j$ depends on the weights $W_{i,j}$ and biases $b_j$ and $d_i$ [see Fig.~\ref{fig:experimental_setup}(c)], which are summarized as the set of variational parameters $\mathcal{W}$. $Z\left(\mathcal{W}\right)=\sum_{\left\{\vec{a},\vec{h}\right\}}\mathrm{exp}\left[-E\left(\vec{a},\vec{h};\mathcal{W}\right)\right]$ ensures normalization. 
Here we consider specifically the tetrahedral and Pauli-4 POVMs, which have the minimal amount of four possible measurement outcomes, so the multinomial visible units $a_i$ can take four different values. We use a one-hot encoding of their states \cite{Carrasquilla2019}.
The model distribution $Q(\vec{a};\mathcal{W})$ is fitted to the measured distribution of outcomes, which follows $P(\vec{a})$, by contrastive divergence learning \cite{Hinton2012}. In this approach, the Kullback-Leibler (KL) divergence from the model to the target (data) distribution,
\begin{equation}
    D_{\mathrm{KL}}\left(P\|Q\right)=\sum_{\left\{\vec{a}\right\}}P\left(\vec{a}\right)\mathrm{ln}\left[\frac{P\left(\vec{a}\right)}{Q\left(\vec{a};\mathcal{W}\right)}\right]\,,
	\label{eq:DKL}
\end{equation}
is considered.
This divergence can be interpreted as a measure for the distance between the two distributions and hence the model distribution can be optimized to approximate the target distribution by minimizing $D_{\mathrm{KL}}\left(P\|Q\right)$. The optimization is done by updating the model parameters via gradient descent, following the negative gradient of the KL divergence.

For training we adopted the numerical routines provided along with Ref.~\cite{Carrasquilla2019}. We refer to this method of direct fitting of the POVM probability distribution as the \emph{POVM ansatz} in the following.

Positivity of the density matrix can be enforced by using the latent space purification ansatz \cite{Torlai2018a}. This ansatz exploits the fact that any density matrix can be re-expressed as the partial trace over a pure system with twice as many degrees of freedom. Mathematically
\begin{equation}
    \rho_{\svec{\sigma}\svec{\sigma'} } = \sum_{\svec{\alpha}} \Psi^*_{\svec{\sigma} \svec{\alpha}} \Psi_{\svec{\sigma'} \svec{\alpha}}
\end{equation}
where 
\begin{equation}
    \ket{\Psi} = \sum_{\svec{\sigma} \svec{\alpha}} \Psi_{\svec{\sigma} \svec{\alpha}} \ket{\svec{\sigma}}\otimes \ket{\svec{\alpha}}
\end{equation}
is the pure state of system qubits $\svec{\sigma}=(\sigma_1\ldots \sigma_N)$ and ancillary qubits $\svec{\alpha}=(\alpha_1\ldots \alpha_N)$. The coefficients of this wave function in the given reference basis ($\ket{\svec{\sigma}}\otimes \ket{\svec{\alpha}}\in\{\ket{\su_z},\ket{\sd_z}\}^{\otimes 2N}$) can again be parameterized by an RBM-inspired variational ansatz. But note that the wave function coefficients are complex, which requires the RBM ansatz to have complex parameters. Alternatively, one can use one RBM for the encoding of the phase and one for encoding the modulus of the wave function coefficients. We used the latter approach, following Ref.~\cite{Torlai2018}. Training of this ansatz is based on measurements in different bases. In our example of the Pauli POVM, we measured the state in all $9$ possible choices of measuring each qubit in one of the three Pauli-bases. The training algorithm minimizes the sum of the KL divergences between the measured probabilities and the corresponding model prediction over all measured bases. We will refer to this NQS tomography method as the \emph{purification ansatz} in the following.

If system and ancillary system are in a product state, the density matrix defined as the trace over the ancillary system will be pure. Thus, for restricting the variational manifold to pure states, one can simply remove all interactions between system and ancillary qubits in the variational ansatz. For learning the parameters of this \emph{ pure-state ansatz}, again, measurements in different bases are needed \cite{Torlai2018}, for which we again use the Pauli POVM data.

Further assuming that the wave function $\ket{\psi}=\sum_{\svec{\sigma}} \psi_{\svec{\sigma}} \ket{\svec{\sigma} }$ of the state has real non-negative coefficients $\psi_{\svec{\sigma}}$, one can simply parameterize these coefficients by a real RBM \cite{Torlai2019}. In this case a measurement in the computational basis $\{\ket{\svec{\sigma}}\}_{\svec{\sigma}}=\{\ket{\su_z},\ket{\sd_z}\}^{\otimes N}$ is sufficient for training.
We trained the purification ansatz and the pure state and \emph{positive-real wave function ansatz} using the open source library QuCumber \cite{qucumber_citation}.

\section{Two-photon experiment}
\label{sec:experiment}

\begin{figure}
\includegraphics[width=\columnwidth]{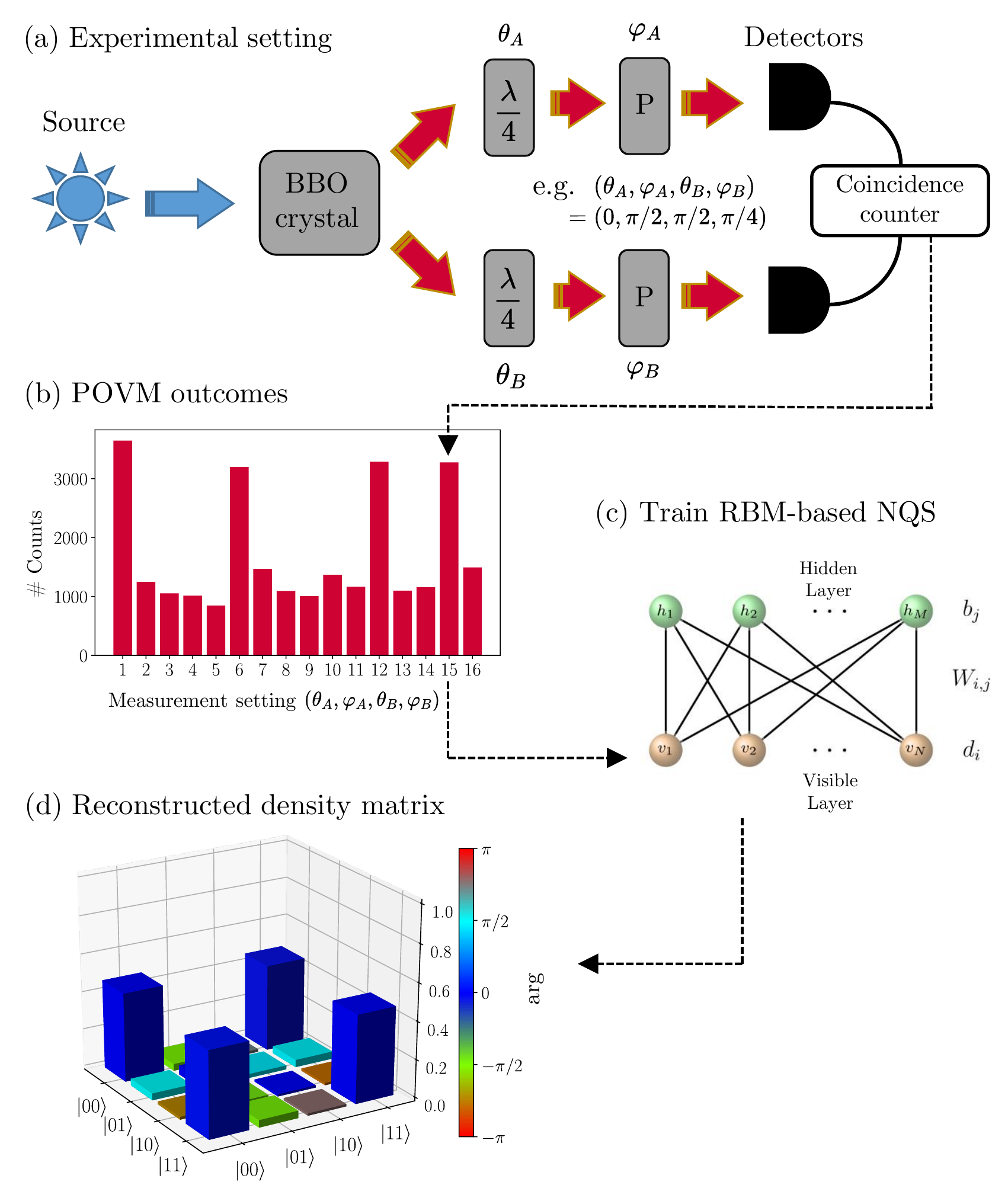}
\caption{Neural network quantum state tomography scheme: (a) Entangled photonic Bell pairs from an SPDC process are measured by coincidence count rates. Arbitrary single-qubit projectors can be selected by adjusting the angles $\theta_A, \varphi_A, \theta_B, \varphi_B$ in $\frac{\lambda}{4}$-plates and polarizers $P$. (b) This is repeated for all elements of the POVM $\{M_{\vec{a}}\}_\vec{a}$ yielding a probability distribution $P(\vec{a})$ describing the state. (c) This forms the training data for different neural network quantum state architectures based on restricted Boltzmann machines (figure adapted from \cite{czischekneural}). (d) After training, the network-encoded states are used to reconstruct the density matrix or to predict observables.}
\label{fig:experimental_setup}
\end{figure}

Our experimental testbed is a commercial photon source \cite{qutools_manual} that produces polarization-entangled photon pairs through type-1 SPDC in a non-linear BBO crystal. A pump laser of $404\,$nm impinges with a linear polarization in a $45^{\circ}$ angle to the optical axes of the crystal, creating superpositions close to the maximally entangled Bell state $\ket{\psi} = \left(\ket{\su_z\su_z} + \ket{\sd_z\sd_z}\right)/\sqrt{2}$, where $\ket{\su_z}$ and $\ket{\sd_z}$ are represented by vertical and horizontal polarization states of the photons.
Pre- and post compensation crystals ensure the spatial and temporal indistinguishability of the non-linear processes producing $\ket{\su_z\su_z}$ and $\ket{\sd_z\sd_z}$.

The single-photon polarization states are subsequently analyzed by a series of a quarter-wave plate (QWP), a polarizer and a single-photon detector, respectively, as illustrated in Fig.~\ref{fig:experimental_setup}(a). By setting the angles of the QWP and the polarizer, arbitrary outcomes $a_i$ can be selected for each photon. The expectation values of the projection operators $M_\vec{a}=M_{a_1}\otimes M_{a_2}$ are obtained by measuring the rate of coincidental counts at both detectors over a fixed period of time. This is repeated for each of the $36$ ($16$) elements in $\{M_\vec{a}\}_\vec{a}$ of the Pauli-6 (tetrahedral) POVM. Assuming a constant production rate of photon pairs, the numbers of coincidence counts for each polarizer setting normalized by the total number of coincidence counts provides the required probability distribution of outcomes [Fig.~\ref{fig:experimental_setup}(b)] from which the density matrix [Fig.~\ref{fig:experimental_setup}(d)]  can be reconstructed. Measuring each bin of the distribution separately in this fashion is obviously not scalable. In an equivalent scalable scenario, one would beam-split each photon into different paths and select one of the single-photon POVM elements in each path. In this setup, detecting coincidence events and recording in which pair of the detectors the photons were observed, is a way of sampling from the full distribution of POVM outcomes in a single measurement setting. 

We benchmark the reconstructed states on their predictions of the famous CHSH Bell parameter $S$ \cite{Clauser1969}. This observable is measured in a Bell test to demonstrate correlations which cannot be reproduced by classical local hidden variable theories \cite{Bell2004}. In such an experiment two particles (photons) are distributed to two parties who independently perform one of two possible measurements on their respective particle. The two parties evaluate the correlation coefficients $E=(N_{\su\su}+N_{\sd\sd}-N_{\su\sd}-N_{\sd\su})/N_{\rm tot}$ for different measurement settings parameterized by an angle $\theta$. Genuine quantum correlations are certified if the Bell parameter
\begin{equation}
    \label{eq:bell_parameter}
    S\left(\theta\right) = E\left(0, \theta\right) + E \left( 0, -\theta \right) + E \left( 2\theta, \theta \right) - E(2\theta, -\theta)
\end{equation}
exceeds the classical bound of $|S|\leq 2$. 
We compare the results obtained for this observable when predicting it from the states obtained with the different tomography schemes to direct measurements of $S$ as a function of $\theta$.

\section{NQS tomography benchmark}

\begin{figure}
\centering
    \includegraphics[width=\columnwidth]{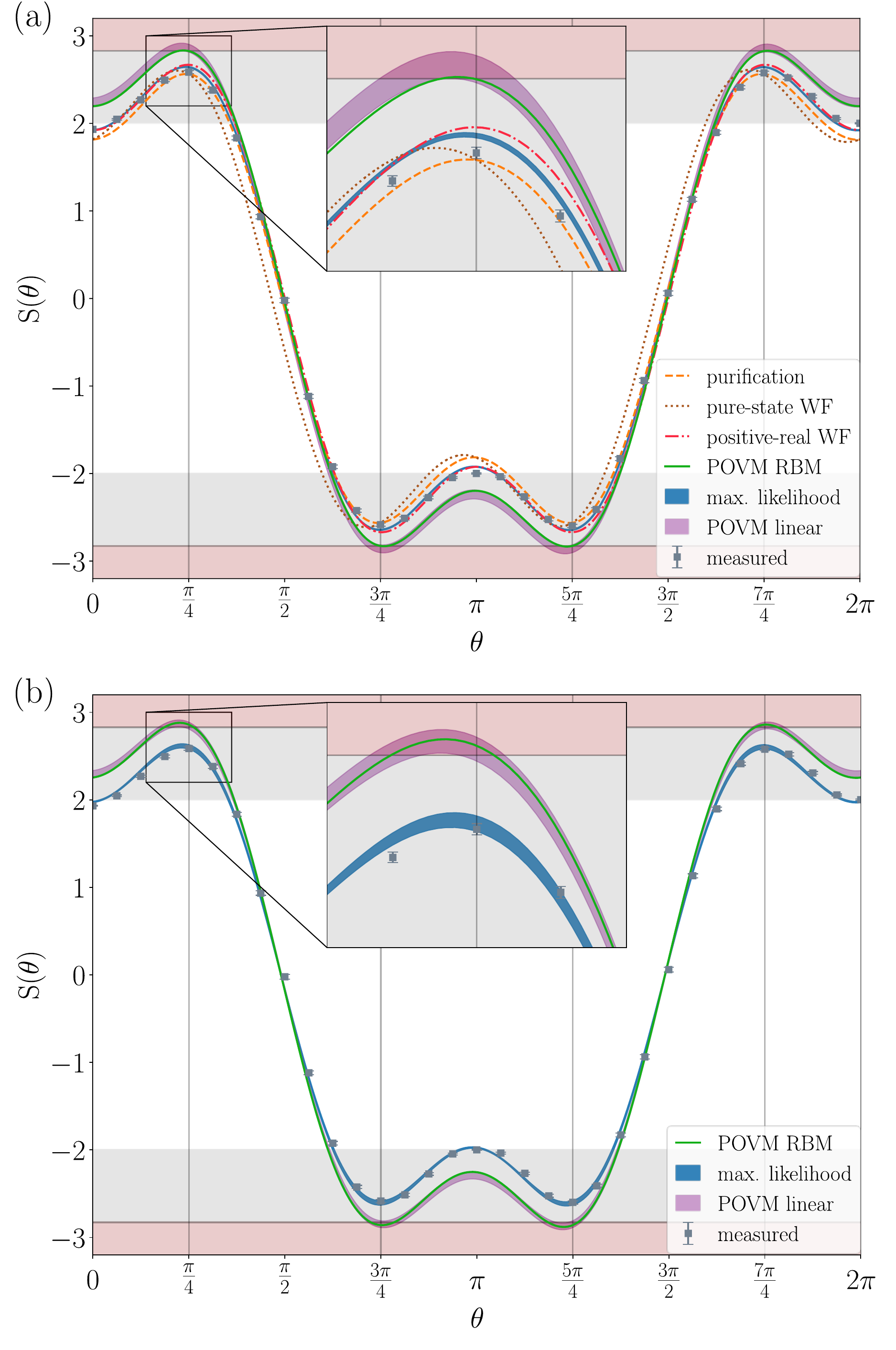}
\caption{Comparison of the predicted Bell parameter from the different NQS tomography schemes to direct measurements and maximum likelihood estimation. (a) Training with measurements in the Pauli product bases applicable to all considered NQS tomography schemes (b) Training with measurements of the tetrahedral POVM applicable to the POVM ansatz only.
Grey regions indicate results forbidden by local realism, red regions denote results forbidden for physical quantum states. Error regions for linear reconstruction and MLE are the standard deviations after repeating the evaluation with data resampled from the measured distribution.}
\label{fig:bell_param}
\end{figure}

Our experimental results are summarized in Fig.~\ref{fig:bell_param}. Panels (a) and (b) show the results of the Pauli POVM measurements and tetrahedral POVM measurements, respectively. 
In both panels of Fig.~\ref{fig:bell_param}, the results of the direct measurement of the Bell parameter for various different values of $\theta$ are shown as blue points.
We counted photons for $t_{\rm{count}}= 3\,$s for each measurements setting resulting in $25000$ counts on average per point.
For the tomographic measurements, we independently measured the probability distributions over the outcomes of the Pauli-6 (and thus also Pauli-4) POVM and the tetrahedral POVM using a total of $60000$ and $27000$ coincidence counts, respectively. As a first check we used these tomographically complete measurements to perform a classical MLE of the density matrix. From the resulting density matrix the Bell parameter can again be calculated. The result, shown by the blue line, agrees well with the Bell measurements (except for small deviations close to the extremal points, see inset of Fig.~\ref{fig:bell_param}(a), which may be due to known shortcomings of MLE \cite{Schwemmer2015, Ferrie2018}). This confirms that the experimental conditions were largely stable between different data sets.
We also show the result of a direct linear reconstruction of $\rho$ from the measured normalized count rates according to Eq.~\eqref{eq:lin_rec}, including the statistical error resulting from the finite number of measurements (purple shadings in Fig.~\ref{fig:bell_param}). The Bell parameter obtained for the linearly reconstructed $\rho$ deviates from the directly measured data and even exceeds the value of $S=2\sqrt{2}$ maximally allowed quantum mechanically, which indicates that the reconstruction yields an unphysical density matrix. Indeed one obtains negative eigenvalues for the reconstructed $\rho$ ($\{0.985, 0.078,  -0.007, -0.055\}$ for the Pauli-4 POVM and $\{0.972, 0.11, -0.022, -0.056\}$ for the tetrahedral POVM).

Turning to NQS tomography, starting with the POVM ansatz, we trained a multinomial RBM with three hidden neurons to approximate both the measured Pauli-4 POVM distribution [Fig.~\ref{fig:bell_param}(a)] and the tetrahedral POVM distribution [Fig.~\ref{fig:bell_param}(b)]. In both cases the distribution was learned well by the network. The KL divergence decreased to $\sim 10^{-4}$. This is expected as the number of parameters in the network exceeds the number of bins in the probability distribution, thus overparameterizing the state. The predicted Bell parameter (solid green lines in Fig.~\ref{fig:bell_param}) agrees with the result of the linear reconstruction and clearly deviates from the directly measured values. 
This shows that the linear reconstruction method on which the POVM ansatz \cite{Carrasquilla2019} is based can give unreliable predictions for observables under imperfect measurements and noise, due to the absence of the positivity constraint on the density matrix.

In the case of the Pauli measurement we can also use the obtained data for training the purification ansatz, which does include the positivity constraint. Here we used a network architecture with two neurons for the latent and three neurons in the hidden layer \cite{Torlai2018a}.  This yields a prediction of comparable quality to the MLE, see dashed yellow line in Fig.~\ref{fig:bell_param}(a). This confirms that the constraint of positivity can greatly improve the capabilities of tomography methods for estimating physical observables. 

We now restrict the ansatz to encode only pure states by omitting couplings to the ancillary spins (pure-state ansatz). Although the MLE density matrix has a purity of only $0.91$ for the Pauli and $0.87$ for the tetrahedral POVM measurement this gives a reasonable estimate of the Bell parameter, as indicated by the dotted brown line in Fig.~\ref{fig:bell_param}(a). However, we observe a shift in $\theta$ of the predicted curve with respect to the directly measured Bell correlations. 

If we further simplify the ansatz by assuming real non-negative wave function coefficients, the agreement between prediction (dot-dashed red curve) and data is surprisingly good. This has two reasons: First, the prepared state has small complex phases of the coherences between the components of the Bell state. Second, this reconstruction technique only uses the information of the Pauli $z$-basis measurement, which happens to yield better predictions for the Bell parameter than methods using the full data set. Thus, the good agreement with the directly measured Bell parameter is rather a coincidence and may not hold for other observables as we further discuss below.
We also observed that the fidelity to a perfect Bell state $\ket{\psi} = \left(\ket{\su_z\su_z} + \ket{\sd_z\sd_z}\right)/\sqrt{2}$ is highest for the state reconstructed assuming a pure state with positive-real coefficients. Given that the MLE reconstruction can be regarded as the most reliable estimation of the density matrix, we conclude that the pure state assumption leads to a crude overestimation of the Bell-state fidelity. This shows that making unjustified assumptions about the state can lead to false conclusions about the quality of the prepared state.

\begin{figure}
\centering
\includegraphics[width=\columnwidth]{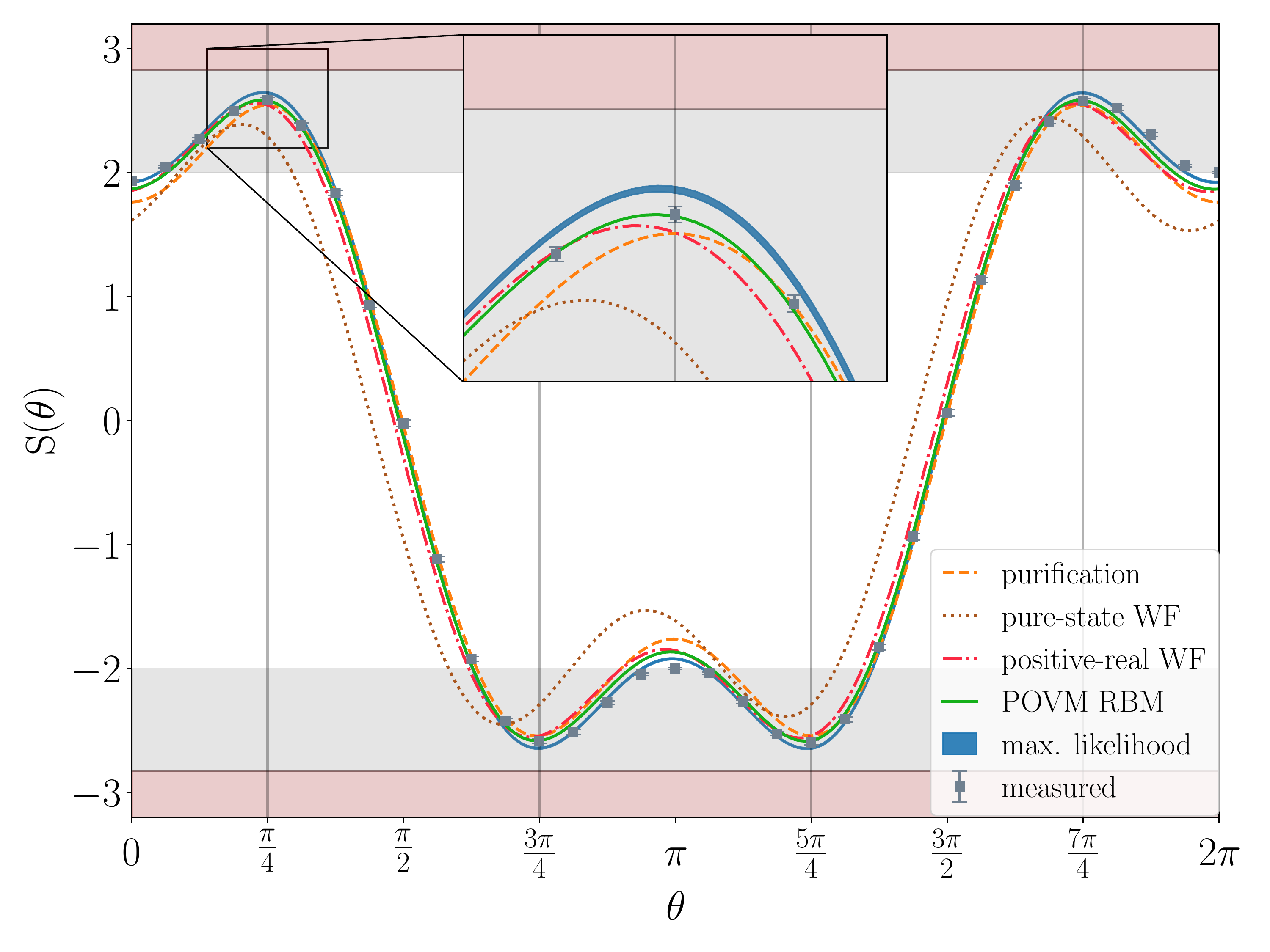}
\caption{Bell parameter estimations when applying the NQS tomography schemes to synthetic data in the Pauli bases sampled from the maximum likelihood density matrix from Fig.~\ref{fig:bell_param}(a). }
\label{fig:synthetic_data}
\end{figure}

We further test our hypothesis that the failure of the POVM ansatz, i.e.\ the linear reconstruction method, is mainly caused by imperfect measurements. For this, we generated a synthetic data set by exact sampling from the outcome distribution predicted by the reconstructed MLE density matrix. Figure~\ref{fig:synthetic_data} shows the results of this procedure for the Pauli measurement. When learning on this synthetic data set all methods perform well confirming that the failure of the linear reconstruction is indeed due to noisy measurements, i.e.\ imprecise settings of the chosen detection angles. The shift in $\theta$, observed for the pure-state ansatz with complex wave function coefficients persists. This is understood by observing that during training, the pure-state ansatz faces an impossible task: The MLE density matrix exhibits considerable mixing, so the pure state will not be able to adapt to the sampled training data’s distribution arbitrarily well. The network instead tries to find a compromise that distributes the inevitable error over all possible bases. The Bell parameter is only sensitive to bases involving the $z$ or $x$ direction, which for the complex wave function training show a considerable error in an attempt to match the data in other bases that are irrelevant to the Bell parameter. In contrast, in the case of the positive-real state the training is only based on $z$-basis data which yields a much better agreement with the independently measured Bell parameter.

\begin{figure}
\centering
\includegraphics[width=\columnwidth]{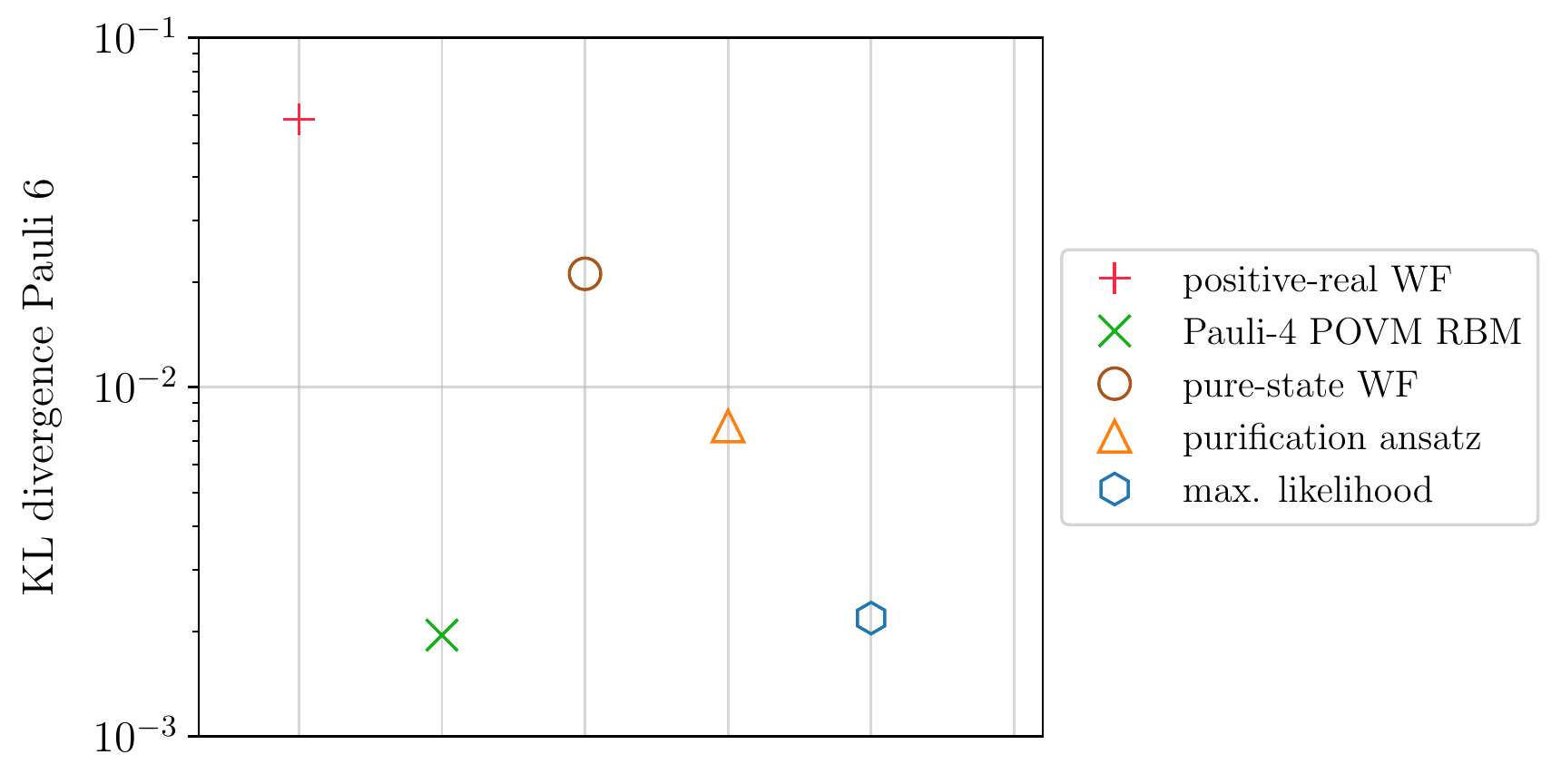}
\caption{Average KL divergence in the Pauli bases between measurements and reconstructed states for all employed tomography schemes. The complexity of the methods increases left to right, expressing the scaling  of computational resources consumed (see text for details).}
\label{fig:DKL}
\end{figure}

Having benchmarked the quality of the different NQS tomography methods by their ability to predict a specific observable, namely the Bell parameter $S(\theta)$, we now analyze their ability to approximate the measured data as a more generic performance measure. We study how well they approximate the data in terms of the loss function that is minimized during the training of the purification ansatz (and of the pure-state and positive-real wave function ansatz), i.e.\ the average KL divergence between measured and predicted outcome distributions of the Pauli-6 measurement, see Fig.~\ref{fig:DKL}.

The quality of the pure-state and positive-real wave function ansatz is clearly worse in this measure than estimations of the general density matrix.
This is expected as the pure-state assumption is not justified. In the case of the positive-real wave function approach only part of the data, namely the measurements in the $z$-basis are used for training resulting in a poor KL divergence to the full data set. This strengthens our conclusion that the good results of the positive-real wave function approach in reproducing the measured Bell parameter are restricted to this specific observable.
The purification ansatz yields results that are not quite as good as the linear reconstruction Pauli-4 RBM and the MLE, cf.\ Fig~\ref{fig:DKL}. This shows that the purification network does not achieve optimal training of its cost function. This could either stem from difficulties of the numerical optimization or from insufficient representational power of the network. The POVM RBM learns the measured Pauli-4 distribution to high accuracy as the learning is not hampered by the positivity constraint. However, it does not resolve some of the details of the full Pauli-6 distribution as some Pauli-6 outcomes are summarized into one Pauli-4 outcome (Eq.~\ref{eq:pauli4_lastline}) causing a residual KL divergence to the full data set. We emphasize that the good performance of the POVM ansatz in this measure does not mean that it performs well in predicting physical observables, as for this task the positivity constraint on the density matrix can be crucial as discussed above.

We sorted the different approaches according to their complexity, referring to the computational resources required for the optimization, or training procedure. The positive-real wave function and POVM approaches involve only models with real-valued parameters and require no unitary rotations into different bases, which makes optimization easier. The pure-state and the purification ansatz were more challenging to train due to the unconventional network architecture required to encode the complex wave function coefficients. This is in contrast to the POVM ansatz where the POVM probability distribution can be learned using a standard RBM architecture. This is the price one has to pay for including the positivity constraint. The maximum likelihood estimate is computationally easier to obtain in this regime of few qubits but scales exponentially for larger systems, which is why we classified it as the most complex method.

\section{Conclusions}

In summary, we have found that including the positivity of the density matrix as a constraint of the variational ansatz in efficient quantum state tomography methods greatly improves the quality of the reconstructed quantum state in terms of predicting physical observables. This may be expected as the constraint results in a smaller set of variational states among which the optimization is carried out. But interestingly, the constraint is not as important for training on synthetically generated data.
From this we conclude that when applying efficient NQS tomography methods to experimental data, imperfections in the measurements can lead to subtle issues which require a careful choice of the state parameterization.
On the other hand, the inclusion of the positivity constraint renders the learning procedure more cumbersome as the purification ansatz imposes a complicated, rather non-standard network architecture.
Making additional assumptions about the state, which are not \emph{a priori} fulfilled, can strongly bias observables predicted from the reconstructed state.

Questions to be addressed in the future concern the scaling properties of neural network state tomography approaches. It has been shown that neural network quantum states can approximate large classes of states efficiently \cite{Carleo2017, Deng2017, Gao2017, Cai2018, Carleo2018, Levine2019, melko2019restricted} with favorable generalization properties \cite{Sehayek2019, Westerhout2020}. Our work shows that for real experimental data, there may be a trade-off between the complexity of the ansatz and the quality of the learned state representations. This is in particular true for the inclusion of the constraint of the positivity of the density matrix, which, in general, leads to an exponential hardness of the optimization problem \cite{Torlai2018a} but also significantly improves the quality of the tomographic reconstruction. It will be crucial to understand whether these issues persist when going to larger system sizes, where the measured distributions are necessarily undersampled and cannot be approximated precisely. In a regime where the generative model only needs to capture the coarse "features" of the quantum state, intuition from machine learning suggests that it may actually be expected to perform better.
Furthermore, it has recently been shown that, in a regime where the Hilbert space dimension is much larger than the number of available measurements, adding the positivity constraint can deteriorate the quality of the reconstructed state \cite{Huang2020, Struchalin2020}. Whether such effects also appear in tomography procedures based on artificial neural networks used in this work will be investigated when scaling up system sizes.
Regarding representational power and learning performance, the choice of the RBM architecture is certainly not without alternative and we will employ and benchmark other generative models such as autoregressive models, autoencoders, or generative adversarial networks in the future.

\acknowledgements
This work is supported by the Deutsche Forschungsgemeinschaft (DFG, German Research Foundation) under Germany's Excellence Strategy EXC 2181/1 -- 390900948 (the Heidelberg STRUCTURES Excellence Cluster), by the DFG -- project-ID 273811115 -- SFB 1225 (ISOQUANT), and by the European Union Horizon 2020 program under grant agreement No. 817482 (PASQuanS).


%

\end{document}